\documentclass[12pt]{article}
\usepackage[usenames]{color} %usado para el color del texto
\usepackage{amssymb} %maths
\usepackage{amsmath} %maths
\usepackage[utf8]{inputenc} %útil para escribir directamente caracteres acentuados
\usepackage[T1]{fontenc}
\usepackage[scaled]{helvet}

\usepackage{sansmath}
\sansmath

\usepackage{lettrine} 
\usepackage[french]{babel} 
\renewcommand\LettrineFontHook{\sffamily\fontseries{m}} 
\renewcommand\LettrineTextFont{\sffamily\scshape}

\begin{document}

\begin{center}
    {\bf \LARGE   Star Formation from Spitzer (Lyman) to Spitzer (Space
Telescope) and Beyond             }\\[4ex]
    {\Large     A summary of Symposium 9, JENAM 2008}\\[4ex]
\end{center}
\parskip=1ex \parindent=3ex     

\begin{quote} 
  {\sl held in Vienna, 10-12 September 2008, and convened by Jo\~ao
    Alves (Calar Alto Observatory) and Virginia Trimble (U. California
    Irvine and Las Cumbres Observatory)}
\end{quote}

\vspace{0.5cm} 

The confluence of the 400th
anniversary of astronomical telescopes, the completion of the basic,
cold, 5-year mission of the Spitzer Space Telescope, and the
near-certain advent of JWST, ALMA, and extremely large, ground-based
telescopes seemed to invite a symposium to investigate the past,
present, and future of star formation studies. While this summary
attempts to mention everybody, with at least one significant idea from
each speaker, including the one-minute poster presentations, it will
surely fail. The sessions were expertly chaired by L. Woltjer,
C. Cesarsky (also involved in the ESO event), J. Andersen, and
H.-M. Maitzen.
 
The Symposium started with two historical introductions (V. Trimble \&
B.G. Elmegreen), addressing, first, the very long time required for
astronomers all to agree, only after 1950, that star formation is an
on-going process, not something that happened long ago (whether
10$^7$, 10$^{10}$, or 10$^{12}$ years ago) when the universe was very
different, and second, the vital roles of Lyman Spitzer, his immediate
predecessors, colleagues, and students, in establishing the existence
and properties of interstellar matter, from which stars could form,
and the processes that would allow them to do so. Remarkably, Spitzer
was never interested in the idea of cold molecular hydrogen as the raw
material of star formation and came rather late to the idea of
turbulence as an important process.
 
We follow the ``seven simplest lessons from 60 years of star
formation'', as outlined by J. Alves, as a logical order to this
summary, and invite you to keep an eye out for some of the topics of
on-going dispute, including (a) whether the initial mass function
(IMF) is universal, what determines it, and whether it is closely
related to the mass distribution of dense cores in pre-stellar clouds
(Core Mass Function or CMF), (b) whether triggering is important, (c)
whether massive stars form the same way as ones that can remain below
Eddington luminosity throughout the process, (d) environmental effects
and the role of binaries, (e) how brown dwarfs form, and (f) how
(in)efficient is star formation, and why. And so on to the seven
``certainties'', keeping in mind that Z is metallicity and z is
redshift.

\begin{enumerate}

\item {\bf Stars form continually in the cold interiors of dark
    molecular clouds} (if you doubt this, please leave the
  room). Multiwavelength studies of specific regions persuaded us all
  to remain (I. Zinchenko, on S76E, with triggering by H$_{\rm II}$
  expansion; M. Rengel on the second class 0 source in Lupus 3,
  indicating these live only 10$^4$ yr; P. Persi on a new SF site NGC
  6334 IV (MM3); and Nakajima, also on the Lupus 3 region).

\item {\bf Star formation is inefficient}, meaning that, if you look
  at a particular mass of cool, dense molecular-gas, the fraction of
  it turned into stars in a dynamical time is typically a few percent
  (J. Silk), though larger values are possible in bound clouds
  (I. Bonnell) and very different numbers probably describe star
  formation in galaxies very unlike the Milky Way and at large z
  (E. Grebel).

\item {\bf Most stars form in groups of 10 - 10$^6$}. Cluster
  environments can enhance disk accretion onto planetary cores
  (S. Pfalzner). Brown dwarfs are more spread out than stars
  (S. Schmeja), though, like the evidence for mass segregation as
  clusters age, this surely has some contribution from source
  confusion in dense centers. 

\item {\bf There is a characteristic product}, a log-normal IMF
  peaking at 0.2-0.3 M$_\odot$ though this too could have been very
  different long ago and far away (Grebel).  Also, low mass stars are
  single (R. Jayawardhana on Cha I and Upper Sco, also providing a
  candidate for the first directly-imaged exoplanet), in contrast to
  Herbig AeBe stars, most of which are binaries, their disks aligned
  with their orbit planes (R. Ooudmaijer).

\item {\bf Feedback processes are ubiquitous and important}. There are
  jets at all wavelengths (K. Stapelfeldt on numerous new Herbig-Haro
  objected detected by SST), the need for ongoing supernovae to keep
  star formation down to the observed 2\% (J. Silk), and perhaps even
  massive star feedback to form clusters (J. Alves).

\item {\bf Stars form with and from accretion disks across the full
    mass range} from BDs to OBs, and there is a definite time sequence
  over which the disks disappear (I. Tsukugoshi on T Tauri
  stars). There are also evolutionary sequences in maser type, radio
  emission, and SED shapes (R. Oudmaijer). Whole clusters also evolve
  (S. Schmeja) from hierarchical to centrally condensed structures.

\item {\bf Nature does some ``pre-packaging''}, so that the
  distribution of core masses, the CMF, has the same shape as the IMF
  (though shifted to larger masses) and must somehow give rise
  directly to the IMF (J. Alves). This was perhaps the topic of
  greatest dispute among the ``certainties''. Several speakers asked
  whether the CMF predicts the IMF (R. Kawabe reporting several
  AzTEC/ASTE surveys; R. Smith noting that different methods yield
  different observed CMFs; P. Hennebelle remarking on the range of
  relevant processes, with outflows, accretion, and turbulence of
  comparable importance; and S. Dib suggesting that the transformation
  from CMF to IMF is a function of environment), I. Bonnell firmly
  denied a directly link between CMF and IMF once one allows for
  continuing fragmentation as well as core0 accretion.

\end{enumerate}

Not yet at the level of eternal verities are the primacy of massive
stars in the formation process (with disk accretion, competitive
accretion, and stellar collisions and mergers in environments of
increasing density, according to R. Klein, and the private opinion of
VT) the need for all the processes you can think of (gravity, angular
momentum transfer, magnetic fields, accretion, turbulence, feedback -
this is either the good news or the bad news, depending on how you
feel about programming). But the probability that there is no further
missing physics counts as good news. 

Then came four outstanding review talks, two from observers two from
theorists (and if you are organizing a seminar series this year, try
to get at least one of these speakers!). First, K. Stapelfeldt
provided an overview of the Spitzer mission, the five-year cold part
of which is essentially over, but a two-year ``warm'' extension, during
which the two shorter wavelengths will still be usable, has been
approved. SST is currently about 1 AU from Earth, drifting backwards,
and eventually will not be able to turn in the right direction and
send us data. 

Among the discoveries important for star formation have been,
\begin{itemize}
\item 70\% of infrared dark clouds have embedded protostars (and those
  that do not could have BDs or might eventually disrupt)
\item at least one region has remarkably gray dust with
  A$_{24\mu\mathrm{m}}$/A$_{\rm K}$ $=$0.44 there is spectroscopic evidence
  for many kinds of grains, including large ice-mantled ones
\item water is found in many places as vapor or ice; there is also
  acetylene
\item the statistics of class 0, I and II sources are not quite as
  expected
\item disks with central holes, perhaps due to planets, are fairly
  common
\item protostellar disks last 10$^7$ years and debris disks 10$^8$
  years; debris disks imply that agglomeration has proceeded at least
  as far as planetesimals, comets, and asteroids
\end{itemize}

\vspace{0.5cm}

Second, E. Grebel absolutely blasted through the very different
contexts in which star formation occurs, from starbursts down to dwarf
galaxies, pointing out the different rates, patterns, efficiencies,
and probably IMFs, and the evidence for different modes in common
galaxy types, as observed or as inferred from the resultant star
populations. Continuous, episodic, or one-shot star formation occurs
depending on gas content, mass density of the galaxy, and interactions
or accretion. Some other points she made (far from a complete list)
include, 

\begin{itemize}
\item stars are now forming in S and Irr galaxies, in galactic
  centers, and in interacting galaxies. Star bursts process 100
  M$_\odot$/yr and ULIRGs up to 1000 M$_\odot$/yr
\item typical spirals form 20 M$_\odot$/yr, much larger than the Milky
  Way value of 1-3 M$_\odot$/yr
\item for many gE's the rate is roughly 0 M$_\odot$/yr, but about 1/3
  have evidence (including Galex UV colors) for active rather than
  passive evolution, that is some on-going star formation
\item field gE's have their oldest stars about 2 Gyr younger than cluster
  gE's
\item E+A galaxies indicate cessation of star formation at a definite time
  in the past
\item the Milky Way has a number of discrete stellar populations,
  distinguished by age and Z, including globular clusters (not
  themselves all the same) two sorts of field halo stars, two sorts of
  disk stars, and a bulge
\item there was a time gap between the end of halo and beginning of
  disk star formation in the MW which is not understood; the bulge
  stars are mostly older than 10 Gyr and have [Fe/H] across the range
  -2.0 to +0.5

\item most large galaxies show age and metallicity gradient

\item it is not clear whether Irr galaxies have massive halos; the star
  velocity dispersion is close to rotation speed, and HI tends to be
  spherical (consider maps of LMC)

\item IR galaxies host 10-20\% of current star formation

\item there are tidal tail galaxies and BCDs (with HI and star formation
  concentrated at their centers)

\item dwarf galaxy SF is very inhomogeneous, and you can see pollution by
  single SNe as scatter in relative abundances

\item the ratio of s to r products is an age indicator
\item winds are important

\item star formation in the outskirts of S's is not understood

\end{itemize}

\vspace{0.5cm}

Third, J. Silk described the multitude of physical processes that must
be considered in theories of star formation, the evidence for them,
and some of the outstanding questions. Key issues include the IMF,
star formation efficiency, turbulence, quenching, and
triggering. Among the points he made were,

\begin{itemize}
\item the IMF is not necessarily constant, and if it was top heavy at
  large z, this will affect the SFR(z) you derive from any tracer

\item the mass assembly history derived from SST and star formation
  histories derived by other methods disagree at z$=3-4$; differences
  in stellar M/L (the IMF) are a likely cause

\item core velocities are mildly supersonic in the $\rho$ Oph region; more
  generally, porosity of the ISM is self-regulated, so that star
  bursts have high turbulence and low porosity, while quenching occurs
  with low turbulence and high porosity

\item the percentage of gas in GMCs is also regulated by turbulence

\item quenching is due to different processes on different scales and
  in galaxies of different masses, for instance fountain and outflows
  on large scales in normal galaxies, but BH accretion, jets, and
  radiation in AGNs, whose activity is quenched at the same time,
  corresponding to the well-known black hole$-$bulge relation
\item triggering is seen on
assorted scales but is not universal
\item AGNs can also enhance star
formation by compressing gas, and the SFR depends on interactions
between hot and cold gas
\item downsizing means both that big halos formed
first and that the ratio of (SFR)/M(already in stars) declines toward
the present from z=2.5. The process is perhaps magnetically regulated.
\end{itemize}

\vspace{0.5cm}

Fourth, the primary discussion of star formation calculated from
numerical simulations came from I. Bonnell, for whom the key questions
are the why's of star masses and the IMF, of inefficiency, of clusters
vs. distributed SF, and the how of core properties giving rise to star
masses. On this last point, he firmly concluded that, because of
on-going accretion plus fragmentation, it is very unlikely that there
is a 1:1 relation between core mass and stellar mass. Initial
conditions are obviously important for these simulations, so that the
SST survey of GMCs (the stage where $\rho =$10$^{-17\dots -21}$ g/cm$^3$) is
vital input. Other things that matter include binaries and disks. Most
star formation occurs in bound structures, where low mass stars and
BDs form from gas falling into the cluster, while high mass stars
result from rapid accretion (slowed but not stopped by feedback) in
incipient cluster cores. Bound gas clouds have SFE around 15\% vs. 3\%
for unbound clouds.

Several of the shorter contributions were of direct relevance to these
issues, for instance high resolution mapping of Av in Barnard 59 as a
probe of SF efficiency (C. Roman), the need (in calculations) for
external confining pressure to keep gas together and allow small
length-scale fluctuations to grow (J. Dale), the dominance of small
separations and mass ratios near one for low-mass binaries
(R. Jayawardhana), and the significantly larger luminosities of
ultracompact Hii regions compared to massive YSOs (R. Oudmaijer),

And the future came at the end. We heard about several ongoing and
upcoming projects, including, 

\begin{itemize}
\item the APEX, Atacama Pathfinder, which sees known SF regions,
  starless cores, hot molecular cores, IRAS sources, embedded
  clusters) and CH30H maser sources, for which follow-up searches with
  Effelsberg, IRAM, and Mopra yielded only one non- detection, a
  planetary nebula! (F. Schiller)
\item SOFIA is coming, with a call for proposals due in December 2008
  (M. Hannebush), and more about SOFIA from R. Klein, who pointed out
  that one of its major goals is to identify the dominant formation
  mechanisms for massive stars, though he left the impression that
  everything that anybody has suggested happens somewhere.
\item an all-sky map of Galacic GMCs now in progress, derived from
  2MASS extinction measurements (J. Rowles)
\item a concept study for a 4-meter space telescope usable from mid UV
  to near IR (R. Jansen)
\item a survey of Gould's belt (the diffuse material primarily, not
  the OB star) with HARP on the JCMT; and SCUBA-2 is coming in 2009
  (J. Hatchell)
\item ALMA, for which L. Testi described the science goals, required
  capabilities (in terms of mm/submm resolution of
  0.1$^{\prime\prime}$ and sensitivity sufficient to map CO and [CI]
  over the entire Milky Way), and timeline. But, he said, it will
  neither image exoplanets ``nor solve the star formation problem''
  (partly, one suspects, because it is a little difficult to decide
  just what ``the'' star formation problem is).
\end{itemize}

Our grandest view of the future came from M. McCaughrean who
emphasized the facilities that will become available over the next
decade or two, including ALMA, the large, ground-based E-ELT (plus the
TMT and GMT), radio facilities like e-MERLIN, LOFAR, and SKA, and, in
space, the upgraded HST, Herschel,Sofia, Gaia, and Kepler. But, he
concluded, the most important new facility will be JWST, with a
five-year mission promised and the potential for another five years
before gases and such run out. He indicated that the single most
important thing it has to offer is greatly improved angular resolution
and that, similarly in planning the new, large ground-based
telescopes, the best possible angular resolution is more important
than pushing into the thermal infrared. Goals are
0.01-0.1$^{\prime\prime}$, though one dan make this sound more
impressive by speaking of 10-100 miliarcseconds. Some of these
facilities will return data by the Tera- and PetaByte, so that
improved capacity for number receiving, storing, processing, and
crunching will also be vital. An interesting case (not mentioned) is
LSST, where the decision has to be made just how much raw data can be
kept, so that, for instance, if a flare occurs in a star formation
region somewhere far away, one can go back over the past years'
images, where the source may have been a two-sigma, three photon
smudge, and determine how bright and how variable it was previously.

\vspace{1cm}
Jo\~ao Alves (Calar Alto Observatory) 

\vspace{-0.2cm}
Virginia Trimble (Univ. of California Irvine and Las Cumbres
  Observatory)

\end{document}